\documentclass[pre]{revtex4}
\usepackage{dcolumn}
\usepackage{graphicx}
\usepackage{latexsym}
\usepackage{amsmath}
\usepackage{amssymb}
\usepackage{amsfonts}
\usepackage{amsthm}

\begin{document}

\title{ Phase modulated solitary waves controlled by bottom boundary condition}
\author{ Abhik Mukherjee and M.S. Janaki}
\affiliation{Saha Institute of Nuclear Physics, \\ I/AF, Bidhannagar\\
Kolkata, INDIA}

\begin{abstract}

A forced KdV equation is derived to describe weakly nonlinear, shallow water surface wave propagation over non trivial bottom boundary condition.  We show that different functional forms of bottom boundary conditions self-consistently produce
different forced kdV equations as the evolution equations for the free surface.  Solitary wave solutions  have been analytically obtained where phase gets modulated controlled by bottom boundary condition whereas  amplitude remains constant.
\end{abstract}


\maketitle

\begin{center} {\bf 1. Introduction} \end{center}
The dynamics of the shallow water nonlinear, unidirectional, dispersive, gravity induced surface waves is described by the celebrated KdV equation\cite{kn:kdv}
that admits solitary wave  solutions. The derivation of such an equation assumes that the fluid is incompressible and inviscid, bounded below by a rigid, impermeable bottom and above by a free surface.
 The generalization of the KdV equation to higher order nonlinearities\cite{kn:burde} and multidimensions\cite{kn:infeld} lead to a multitude of
 nonlinear equations that found potential applications\cite{kn:ablowitz} in various physical situations.    The classical KdV equation was
 also extended to admit arbitrary forcing functions leading to the forced KdV equation\cite{kn:Wu}-\cite{kn:orlowski}, subject to  asymptotic initial conditions depending on the forcing disturbances  and also including the effects of surface pressure and topography.  In this work, we also encounter a series of forced KdV equations as the evolution equations of shallow water, nonlinear dispersive waves over non-trivial bottom boundary conditions.  
 The different functional nature  of this fixed bottom condition self-consistently generates different types of forced kdV equations.

 In most of the realistic situations, water waves propagate over a porus bed so that one needs to consider the transformation
of the waves brought about by  the  permeability of the bottom bed. 
{\bf{Several important problems  exist concerning the effects of soil dynamics 
on waves.  Many kinds of structures such as ships, buoys, breakwaters, submersibles
supporting oil drillings rigs are directly or indirectly supported by the bottom bed
that can be made up of a variety of different soils ranging from solid rock, to sand
to fine clay. Persistent attack by the waves causes the varying wave pressure to induce considerable 
stress and strain on the  bottom bed.  This in turn effects the dynamic stability of the bed
causing fatigue of the structures supported by the bed.  A description of the porus bed needs 
different types of constitutive equations depending on the soil type and strain magnitude,
while the fluid medium  is itself described by Darcy's law.  Mei\cite{kn:mei} has developed several theoretical concepts needed to pursue the problems of wave induced stresses in a porus media
using the boundary layer approximation that facilitates even the nonlinear modelling of seabeds.}}

The dynamics of the linear water waves in a channel of permeable bottom has been one of  the  interesting research problems in water wave theory undertaken {\bf from the early times}\cite{kn:Kajiura}-\cite{kn:Hunt}.  Rigorous development of mathematical models\cite{kn:boussinesq}-\cite{kn:boussinesq2} for nonlinear, diffusive, weakly dispersive  water waves interacting with a permeable bottom has begun only in the last decade, with the description based on the Boussinesq approximation.
In shallow water, Boussinesq equation gives wave solutions  propagating in both positive and negative  directions. However, for unidirectional wave propagation in shallow water, the  KdV equation  appears as a reasonable dynamical equation when the vertical fluid velocity at the bottom is assumed to be zero. In this work, we consider the non zero vertical fluid velocity at the bottom that leads to a series of forced KdV equations self-consistently where the functional forms of the leakage velocity appears as  forcing function. 
One of the boundary condition to the problem arises, as is the usual practice,  by considering the upper boundary of the fluid to be a free surface. For the other fixed boundary condition,
a nonzero value of the vertical fluid velocity at the bed of the channel is considered that physically describes  the  existence of a leakage velocity at the interface of the fluid and the bed.
This fixed bottom boundary condition is the key feature in the present  problem.   Situations  describing different types of  interactions of the fluid and porous bed can be analyzed by considering different mathematical forms of this leakage velocity as functions of time, space or both.
The basic features of analysis  of this work run parallel  to the derivation of  KdV equation\cite{kn:Lakshmanan}  in  a hard bottom channel.  The novelty of the present work is that exact solitary wave solutions of the forced KdV equations  have been obtained analytically for different leakage conditions.  For example,  constant, time dependent, space dependent or both space-time dependent  forms of leakage velocity have been considered that control the phase modulation of the obtained solitary wave solutions leading to different types of dynamical behaviour of such waves.

 The arrangement of the paper is as follows.
 In section 2, the foundation  of the problem has been established for a nonlinear shallow water wave propagation using a general  space time dependent
 bottom boundary condition.  For better understanding of the problem, different types of  functional dependence of
 the vertical fluid velocity at the bottom have been considered.  In section 2.1, it is assumed to be constant
 leading to a constantly decelerating solitary wave. In section 2.2, only temporal dependence is considered, which also leads to decelerating solitary wave with the phase modulation explicitly depending on the nature of time dependence.  Both of these two cases have been analytically solved to obtain exact solutions.  
 In the next subsection, spatial dependence has been considered, leading to inhomogeneous KdV, where the inhomogeneity is in the form of a   kdV like equation.  Using the interesting nature of the inhomogeneity,
 exact and approximate solutions are obtained.  In the last subsection, the more general space-time dependent bottom boundary condition is assumed.   In order to obtain exact analytic solitary wave solution, bilinearization technique is applied where also an arbitrary function dependent on time modulates the phase of the solution.  In all the cases considered, the amplitude of the solitary wave remains constant.  Appropriate three dimensional plots of the solution have been presented for each case.


\begin{center} {\bf 2. Derivation of the free surface evolution equation in presence of water leakage at the bottom}\end{center}

 A one dimensional unidirectional wave motion propagating through a shallow  water channel of permeable bottom is considered. The channel is of
uniform cross section and of constant depth $h$. The
 fluid  is assumed to be  incompressible with the wavelength, 
 amplitude and  velocity of the wave represented by l, a  and v respectively (as shown in Figure 1). 
 {\bf The surface tension and viscosity have been neglected throughout this calculation}.
 At
an arbitrary ($x, t$) the free surface displacement is denoted by  $\eta(x,t)$ .
 Two  natural small parameters $\epsilon = {a}/{h}$ and $\delta = {h}/{l}$ are introduced, both of which are $ \ll$ 1, and further $\epsilon$ $ \approx $ $\delta^2  $.

\begin{figure}[!h]
\centering
{\includegraphics[width=7 cm, angle=0]{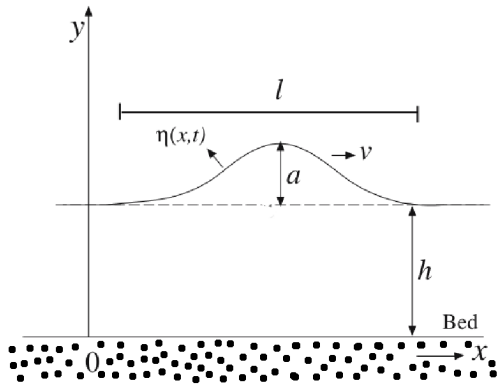}}

\caption{Shallow water solitary wave in a water channel of permeable bottom}
\label{Figure 1}
\end{figure}

The fluid motion can be described by the velocity vector $\vec{V} = V_{h}\vec{i} + V_{v}\vec{j}$ where the subscripts $h,v$ denote horizontal and vertical components of the fluid velocity respectively. From the condition of irrotational flow of the fluid we can
introduce a velocity potential $\phi (x,y,t)$ such that $\vec{V}
 = \vec{\nabla}\phi$.

 Since the fluid is incompressible,  the mass conservation equation leads to
\begin{equation}
 \nabla^2 \phi = \label{poisson} 0
\end{equation}
Again from the momentum conservation equation, i.e, Euler equation, we get,
\begin{equation}
 \frac{\partial \phi}{\partial t} + \frac{1}{2}(\vec{\nabla}\phi)^2 + \frac{p}{\rho} + g y = 0,
\label{Lagrange} \end{equation}
where $\rho, p , g $  are density, pressure of the fluid and acceleration due to gravity
 respectively. {\bf Eq. (\ref{Lagrange}) is well known as the Lagrange equation.}
 {\bf When $\phi$ is independent of time, then the above equation is called Bernoulli's condition.}
  Equations (\ref{poisson}), (\ref{Lagrange}) are the  two main equations
of the problem which must be supplemented by appropriate boundary conditions.

{\bf The fluid is bounded by two surfaces, one is the fixed bottom and other is the free boundary.
Since at the upper free surface, $p=0$, hence taking derivative of eq. (\ref{Lagrange}) along the  direction of propagation, we obtain 
\begin{equation}
 \frac{\partial V_{h}}{\partial t} + V_{h}\frac{\partial V_{h}}{\partial x} + V_{v}\frac{\partial V_{v}}{\partial x} + g \frac{\partial
\eta}{\partial x} = 0
\label{bc1} \end{equation}
Again, at  the free surface
\begin{equation} y(x,t) = h+\eta(x,t) \label{freeb} \end{equation}
Taking time derivative of eq.(\ref{freeb}), we get 
\begin{equation}
 V_{v} = \frac{\partial \eta}{\partial t} + \frac{\partial \eta}{\partial x} V_{h}
\label{bc2} \end{equation}.
These two equations (\ref{bc1}) and (\ref{bc2}) are defined at the free surface of the wave.
Since the upper free surface is movable, hence these equations are called variable boundary conditions.

Since  some amount of water is considered to be leaking through the fixed bottom of the channel, 
 hence the downward vertical fluid velocity at the bottom is nonzero.  This constitutes the fixed 
 boundary condition defined by the equation
 \begin{equation}
 V_{v}(x,0,t) =\frac{\partial \phi(x,0,t)}{\partial y} = C (x,t)
\label{fbc}
\end{equation}
 where $C(x,t)$ is the vertical  fluid velocity at  the bottom of the channel.
This equation is called the penetration condition. }

{\bf Thus ultimately we get  two  equations (\ref{poisson}), (\ref{Lagrange}) that are valid
in the bulk of the fluid.  Taking the derivative of eq. (\ref{Lagrange}) and eq. (\ref{freeb})
at the free boundary, we get
 two    nonlinear boundary conditions (\ref{bc1}), (\ref{bc2}) respectively and  the penetration
condition given by (\ref{fbc}). } 

 The velocity potential $\phi$  is expanded in Taylor series as follows
\begin{equation}
 \phi(x,y,t) = \sum_{n=0}^{\infty} y^n \phi_{n}(x,t)
\label{phit} \end{equation}
Where $\phi_{n}(x,t) = \frac{\partial^n\phi}{\partial y^n}$ at y= 0. Substituting this in the Laplace's equation (\ref{poisson})
the following recurrence relation is obtained
\begin{equation}
 \frac{\partial^2 \phi_{n}}{\partial x^2} = -(n+2)(n+1)\phi_{n+2}
\label{recur} \end{equation}
Using penetration condition ({\ref{fbc}) we can arrive at $\phi_{1}(x,t) = C (x,t) $ and
 using this in the recurrence relation (\ref{recur})  the expression for $\phi(x,y,t)$
is obtained as -
\begin{equation}
 \phi(x,y,t) =  \sum_{m=0}^{\infty}\frac{(-1)^m}{(2m)!} F_{2m} y^{2m} + \sum_{m=0}^{\infty}\frac{(-1)^m}{(2m+1)!} C_{2m} y^{2m+1}
\label{phi} \end{equation}
where $F = \phi_{0}(x,t)$ and the subscript -(2m) in C and F denotes 2m-th order derivative w.r.to x.
The horizontal
and vertical components of the fluid velocity at the free surface are determined as,
\begin{equation}
 V_{h} = y C_{x} - \frac{1}{3!}C_{xxx}y^3+ f - \frac{1}{2!}y^2 f_{xx} + h.o.t
\label{horv}\end{equation}
\begin{equation}
 V_{v} = C -\frac{1}{2!}C_{xx}y^2 - y f_{x} +\frac{1}{3!}y^3 f_{xxx}+ h.o.t
\label{verv} \end{equation}
where  $x$ in the subscript denotes partial derivative with respect to $ x$, $f = \frac{\partial F}{\partial x}$ and h.o.t denotes higher order terms
in $y$.
The different dimensional variables that have made their appearance in the problem  will be made dimensionless by incorporating the following scaling
of variables so that small parameters $\epsilon,\delta$ creep into the equations and
smaller terms can be neglected in comparison to them. 

\vspace{0.5cm}

\begin{center} {\bf {2.1 ~{Scaling of variables}}} \end{center}

All the dependent and independent variables occurring in the above equations are scaled in the following way
by taking account of the smallness parameters $\epsilon,\delta$

$\botfigrule$ $$ x \Rightarrow l x', \eta \Rightarrow a \eta', t \Rightarrow\frac{l}{\sqrt{g h}} t',
 V_{h} \Rightarrow \epsilon \sqrt{g h} V_{h}', V_{v} \Rightarrow \epsilon \delta
\sqrt{g h} V_{v}',
$$

$$f \Rightarrow \epsilon \sqrt{g h} f',
y\Rightarrow h(1 + \epsilon \eta'), C \Rightarrow\epsilon \delta \sqrt{g h} C ',
$$
where the variables in prime are dimensionless and henceforth all terms
 $\backsimeq$ $\epsilon \delta^2$ will be neglected by considering them to be  small compared to terms of the order of $\epsilon$ or $\delta
^2$.
Using this scaling in
equations (\ref{horv}), (\ref{verv}) dimensionless velocity components are obtained  as-

\begin{equation}
 V_{h}' = \delta^2 C'_{x'} + f' - \frac{1}{2}\delta ^2 f'_{x'x'}
\end{equation}
\begin{equation}
 V_{v}' = C' -\frac{1}{2}\delta ^2 C'_{x'x'} - (1 + \epsilon \eta') f'_{x'} + \frac{1}{6}\delta^2 f'_{x'x'x'}
\end{equation}
Hence from the two nonlinear boundary conditions (\ref{bc2}), (\ref{bc1}) we get
\begin{equation}
 \eta'_{t'}+ f'_{x'}+ \epsilon f' \eta_{x'}+ \epsilon \eta' f'_{x'}- \frac{1}{6}\delta^2 f'_{x'x'x'}
= C'- \frac{1}{2}\delta^2 C'_{x'x'}
\label{bc1dl} \end{equation}
\begin{equation}
 \eta'_{x'}+ f'_{t'}+ \epsilon f' f'_{x'} - \frac{1}{2}\delta^2 f'_{x'x't'}
= - \delta^2  C'_{x't'}
\label{bc2dl} \end{equation}
For notational convenience  the prime symbol will  be omitted in all the variables in the subsequent analysis,
 remembering however  that all  variables  correspond  to rescaled quantities.
These are the equations related to the displacement of the free surface wave $\eta$, function related to
velocity potential $f$ and the leakage velocity $C$.

In order to formulate the problem in a  more general way  $ C(x,t)$ is considered  to have different forms.
 In section(2.2), $C$ is considered to be a constant i.e.,
leakage velocity of water at the bottom is constant throughout its motion.

\begin{center} {\bf  2.2 ~C is pure constant} \end{center}
Considering $C$ to be constant, equations (\ref{bc1dl}), (\ref{bc2dl}) can be written as
\begin{equation}
 \eta_{t}+ f_{x}+ \epsilon f \eta_{x}+ \epsilon \eta f_{x}- \frac{1}{6}\delta^2 f_{xxx}
= C
\label{etat}\end{equation}
\begin{equation}
 \eta_{x}+ f_{t}+ \epsilon f f_{x} - \frac{1}{2}\delta^2 f_{xxt}
= 0 \label{etax}
\end{equation}
Expanding $f$ in a  series of small parameters as
\begin{equation}
 f = f^{(0)}+ \epsilon f^{(1)}+ \delta^2 f^{(2)}+ h.o.t
\label{fexp} \end{equation}
and neglecting  higher order terms in $\epsilon$ or $\delta^2$,
 equations (\ref{etat}), (\ref{etax}) converge to -
\begin{equation}
 \eta_{t}+ f^{(0)}_{x}+ \epsilon (f^{(1)} _{x}+  \eta f^{(0)}_{x}+\eta_{x} f^{(0)})
+\delta^2 (f^{(2)}_{x}- \frac{1}{6}f^{(0)}_{xxx})= C
\label{etat2}\end{equation}
\begin{equation}
 \eta_{x}+ f^{(0)}_{t}+ \epsilon (f^{(1)} _{t}+  f^{(0)} f^{(0)}_{x})
+\delta^2 (f^{(2)}_{t}- \frac{1}{2}f^{(0)}_{xxt})= 0
\label{etax2} \end{equation}
In order that equations (\ref{etat2}), (\ref{etax2}) are self-consistent as evolution equations for a one-dimensional
wave propagating along the positive x-axis, the following choice is made:
\begin{equation}
 f^{(0)}= \eta - C t + O(\epsilon \delta^2)
\label{f0} \end{equation}
where $O(\epsilon \delta^2)$ denote terms $\sim$ $\epsilon \delta^2$.
Thus from equations (\ref{etat2}), (\ref{etax2}), we get,
\begin{equation}
\tilde{\eta_{t}}+ \tilde{\eta_{x}}+ \epsilon(f^{(1)}_{x}+ 2 \tilde{\eta}\tilde{\eta_{x}}+
\tilde{\eta_{x}} C t)+ \delta^2(f^{(2)}_{x}- \frac{1}{6}\tilde \eta_{xxx})=0
\label{etat3}\end{equation}
\begin{equation}
\tilde{\eta_{t}}+ \tilde{\eta_{x}}+ \epsilon(f^{(1)}_{t}+  \tilde{\eta}\tilde{\eta_{x}})+
 \delta^2(f^{(2)}_{t}- \frac{1}{2}\tilde \eta_{xxt})=0
\label{etax3} \end{equation}
where $\tilde{\eta} = \eta - C t$.

Let $f^{(1)}, f^{(2)}$ be functions of $\tilde{\eta}$ and its spatial derivatives.
This leads to $f^{(1)}_{t}= -\tilde {\eta_{x}}f^{(1)}_{\tilde{\eta}}+ f^{(1)}_{\tilde{\eta}}
O(\epsilon ,\delta^2)$. where O$(\epsilon ,\delta^2)$ is the term proportional to $\epsilon$ or $\delta^2$.
 Since terms of the order of $\epsilon \delta^2$ are being neglected in the present work, the following relations are obtained

$$f^{(1)}_{t}  \approx -\tilde{\eta_{x}}f_{\tilde{\eta}}^{(1)} =-f^{(1)}_{x}$$
$$f^{(2)}_{t} \approx -\tilde{\eta_{x}}f_{\tilde{\eta}}^{(2)} =-f^{(2)}_{x}$$
$$\tilde {\eta}_{xxt} \approx -\tilde {\eta}_{xxx}$$

Using these results in equations (\ref{etat3}) and (\ref{etax3}), the condition for compatibility of
these  two equations leads to
\begin{equation}
 f^{(1)}_{x} = - \frac{1}{2}\tilde{\eta_{x}}(\tilde{\eta} + C t)
\end{equation}

\begin{equation}
 f^{(2)}_{x} =  \frac{1}{3}\tilde{\eta}_{xxx}
\end{equation}
These results  when substituted into any of the equations (\ref{etat3}), (\ref{etax3}), the following single evolution equation is obtained as
\begin{equation} \tilde{\eta_{t}}+ \tilde{\eta_{x}}+ \epsilon( \frac{3}{2} \tilde{\eta}\tilde{\eta_{x}}+\frac{1}{2}
\tilde{\eta_{x}} C t)+ \delta^2( \frac{1}{6}\tilde \eta_{xxx})=0
\label{kdvcc} \end{equation}

Equation (\ref{kdvcc}) can be converted to a forced KdV equation with a constant forcing term by redefining the
dependent variable as ${\bar \eta }= 3\eta+Ct$.

Equation (\ref{kdvcc}) is similar to KdV equation except for  the 4-th term which comes from the leakage.
 A suitable transformation into a moving frame can remove this term  so that the standard form of 
 KdV equation is recovered.
We use, 
$$ X = x- t - b t^2, T =  t$$
where $b$ is a constant denoting the acceleration of the frame. In the new frame $(X, T)$, equation (\ref{kdvcc}) will look like
\begin{equation}
\tilde{\eta}_{T}+ \epsilon( \frac{3}{2} \tilde{\eta}\tilde{\eta}_{X} + \frac{1}{2}
\tilde{\eta}_{X} C T) + \delta^2( \frac{1}{6}\tilde{\eta}_{XXX})- 2 b T \tilde{\eta}_{X} = 0
\end{equation}
Choosing
$ b = \frac{\epsilon C}{4}$  and
 defining new variables  $u = ({3 \epsilon}/{2 \delta^2}) \tilde{\eta} $, $T'= ({\delta^2}/{6}) T$
the following standard form of KdV equation  is obtained which is in the accelerated frame.
\begin{equation}
u_{T'} + 6 u u_{X} + u_{XXX} = 0
\label{kdvaccs} \end{equation}

{\bf {2.2.1  Solitary wave solution   }}

The well known one-Soliton solution of equation (\ref{kdvaccs}) is given by
\begin{equation}
 u(X,T') = \frac{\beta}{2} Sech^2[\frac{\sqrt \beta}{2}(X - \beta T')]
\end{equation}
where $\beta$ is a constant.
Back boosting the solution to the rest frame we get
\begin{equation}
 u(x,t) = \frac{\beta}{2} Sech^2[\frac{\sqrt \beta}{2}(x - t\{(1 + \frac{\beta \delta^2}{6}) - \frac{\epsilon \mid C \mid}{4} t\})] \label{csol}
\end{equation}
where $\mid C \mid$ is the absolute value of $C$.  The argument of $u(x,t)$
contains linear and quadratic terms in $t$ and the term inside the second bracket behaves like the velocity of the wave.

When $t$ starts increasing from a very small value, the   term  inside the second parentheses decreases 
i.e. the wave retards.

 At a critical time given by
$$ t = t_{c} = \frac{4 (1 + \frac{\beta \delta^2}{6})}{\epsilon \mid C \mid},$$

the second bracketed term vanishes and the wave stops.
After $t_{c}$, the wave propagates in the negative $x$-axis with increasing speed. Since the wave moving in
positive $x$ axis is only of concern to us,  this oppositely moving wave can be neglected in a practical situation.  {\bf Since $\epsilon$ is a small parameter, hence $t_c$ is large for small values of fluid leakage.  For large leakage velocity, $t_c$ become small causing reflection of wave at earlier time}. A x-t plot of the solution (\ref{csol}) is shown in Figure (2) indicating 
that the wave gets reflected at time at $t_c$  and then moves in opposite direction.

\begin{figure}
\centering
{\includegraphics[width=7 cm, angle=0]{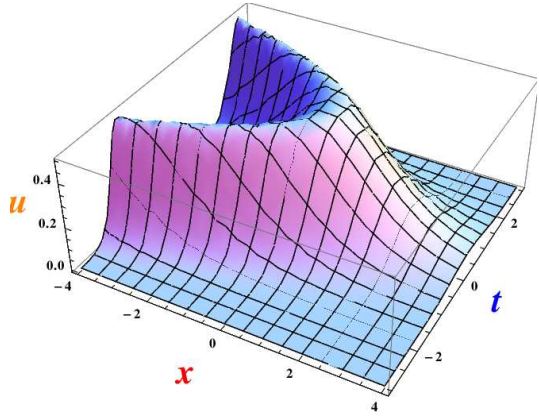}}
\caption{$x-t$ plot of the solution (\ref{csol}) with
   $\beta = 1$, $\mid C \mid= 40000$, $\delta=0.01$, $\epsilon= 0.0001$. $x$ and $t$ are plotted
in the 2 horizontal axes and $u(x,t)$ is plotted in the vertical axis.}
\label{Figure 2 }
\end{figure}

\begin{center}{ \bf 2.3~~ C is function of t only}\end{center}

In the subsection (2.2)  solitary wave solution has been  obtained by considering a constant leakage velocity. The problem is now  generalized  by assuming $C$ to be function of $t$ only.
From equations (\ref{bc1dl}), (\ref{bc2dl})
\begin{equation}
 \eta_{t}+ f_{x}+ \epsilon f \eta_{x}+ \epsilon \eta f_{x}- \frac{1}{6}\delta^2 f_{xxx}
= C (t)
\label{etat4} \end{equation}
\begin{equation}
 \eta_{x}+ f_{t}+ \epsilon f f_{x} - \frac{1}{2}\delta^2 f_{xxt}
= 0 \label{etax4}
\end{equation}
Carrying out series expansion for $f$  as in equation (\ref{fexp}) and neglecting higher order terms in $\epsilon$ or $\delta^2$ the following equations are obtained from equations  (\ref{etat4}) and (\ref{etax4})
\begin{equation}
 \eta_{t}+ f^{(0)}_{x}+ \epsilon (f^{(1)} _{x}+  \eta f^{(0)}_{x}+\eta_{x} f^{(0)})
+\delta^2 (f^{(2)}_{x}- \frac{1}{6}f^{(0)}_{xxx})= C(t)
\label{etat5} \end{equation}
\begin{equation}
 \eta_{x}+ f^{(0)}_{t}+ \epsilon (f^{(1)} _{t}+  f^{(0)} f^{(0)}_{x})
+\delta^2 (f^{(2)}_{t}- \frac{1}{2}f^{(0)}_{xxt})= 0
\label{etax5} \end{equation}
In order to make equations (\ref{etat5}) and (\ref{etax5}) self-consistent, the following choice is made
\begin{equation}
 f^{(0)}= \eta - \int C(t) dt + O(\epsilon \delta^2)
\label{f02} \end{equation}
Thus from equation (\ref{etat5}), (\ref{etax5}) the following equations are obtained -
\begin{equation}
\tilde{\eta_{t}}+ \tilde{\eta_{x}}+ \epsilon(f^{(1)}_{x}+ 2 \tilde{\eta}\tilde{\eta_{x}}+
\tilde{\eta_{x}} B(t))+ \delta^2(f^{(2)}_{x}- \frac{1}{6}\tilde \eta_{xxx}) = 0
\label{etat6}\end{equation}
\begin{equation}
\tilde{\eta_{t}}+ \tilde{\eta_{x}}+ \epsilon(f^{(1)}_{t}+  \tilde{\eta}\tilde{\eta_{x}})+
 \delta^2(f^{(2)}_{t}- \frac{1}{2}\tilde \eta_{xxt}) = 0
\label{etax6} \end{equation}
where $\tilde{\eta} = \eta -  B(t) $ where $ B(t)=\int C(t) dt $.

Considering $f^{(1)}, f^{(2)}$ to be functions of $\tilde{\eta}$ and its spatial derivatives,
$f^{(1)}_{t}= -\tilde {\eta_{x}}f^{(1)}_{\tilde{\eta}}+ f^{(1)}_{\tilde{\eta}}
O(\epsilon ,\delta^2)$. Since terms of the order of $\epsilon \delta^2$ are neglected,
the following relations are obtained
$$ f^{(1)}_{t} \approx -\tilde{\eta_{x}}f_{\tilde{\eta}}^{(1)} =-f^{(1)}_{x}$$ 
$$f^{(2)}_{t} \approx -\tilde{\eta_{x}}f_{\tilde{\eta}}^{(2)} =-f^{(2)}_{x}$$
$$ \tilde {\eta}_{xxt} \approx -\tilde {\eta}_{xxx} $$

Using these results in any of the equations (\ref{etat6}) and (\ref{etax6}) together with  the compatibility condition
leads to the following single equation
\begin{equation}
\tilde{\eta_{t}}+ \tilde{\eta_{x}}+ \epsilon( \frac{3}{2} \tilde{\eta}\tilde{\eta_{x}}+\frac{1}{2}
\tilde{\eta_{x}} B(t))+ \delta^2( \frac{1}{6}\tilde \eta_{xxx})=0
\label{kdvct1} \end{equation}
Equation  (\ref{kdvct1}) can be cast in the form of a forced KdV equation by redefining a new variable as given in the previous subsection with a time dependent forcing term.
An analytical treatment of the influence of the time dependent random external noise on the propagation of nonlinear waves  has been carried out by Orlowski\cite{kn:orlowski} by considering a forced KdV equation. 
Considering the Gaussian character of the noise, the nature of deformation of the stationary solution of KdV-Burgers equation
was studied\cite{kn:wadati}-\cite{kn:zahibo}. during its propagation in randomly excited media.

In order to arrive at a standard form of the KdV equation from equation(\ref{kdvct1}) a transformation to a moving frame given by 
 \begin{equation}  X = x- t - a(t), ~T =  t \end{equation} 
 
 is carried out, where a(t) is a function dependent on $t$.  This leads to 
\begin{equation}
\tilde{\eta}_{T}+ \frac{3 \epsilon}{2} \tilde{\eta}\tilde{\eta}_{X} +  \frac{\delta^2}{6}\tilde{\eta}_{XXX}+
(\frac{\epsilon B(t)}{2} - \frac{\partial a}{\partial t})\tilde{\eta}_{X} = 0
\end{equation}
With the choice ${\partial a}/{\partial t}  = {\epsilon B(t)}/{2}$,
and defining $u = ({3 \epsilon}/{2 \delta^2}) \tilde{\eta} $, $T'= ({\delta^2}/{6}) T$,
we get the standard form of KdV equation in the moving frame
\begin{equation}
u_{T'} + 6 u u_{X} + u_{XXX} = 0
\label{kdvct2} \end{equation}

{\bf 2.3.1 Nature of the Solution }

One-Soliton solution of equation (\ref{kdvct2}) is of the standard form
\begin{equation}
 u(X,T') = \frac{\alpha}{2} Sech^2[\frac{\sqrt \alpha}{2}(X - \alpha T')]
\end{equation}
where $\alpha$ is a constant.  Transforming  this solution to the rest frame 
\begin{equation}
 u(x,t) = \frac{\alpha}{2} Sech^2[\frac{\sqrt\alpha}{2}(x - (1 + \frac{\alpha \delta^2}{6}) t - \frac{\epsilon}{2}\int B(t) dt )]
\label{accsolt} \end{equation}
It should be noted that when $C$ is constant then the integral term inside the argument of $u(x,t)$
is $\frac{ \epsilon C_{abs} t^2}{4}$  which is consistent with the solution of the previous case.

The functional form of  $B (t)$ controls the motion of the solution (\ref{accsolt}). For different choices of the function $B(t)$ the 3D x-t plot of the solution
will have different shapes. If the leakage velocity is dependent on the fluid velocity such that
 when a large upsurge of water arrives in a region,  large leakage occurs
and when t $\longrightarrow$ $\pm \infty$ the leakage goes to zero i.e. leakage is localized in time.  As an example we can choose the functional form of $C$ as, $ C(t) = {b_{0}}/{(b_{1} + b_{2} t^2)}$, where $b_{0}$, $b_{1}$,$b_{2}$ are constants , which  is localized in time.
 The corresponding 3D plot of the solution is given as Figure (3). The solution gets curved due to the presence of nonlinear function of $t$.

\begin{figure}[!h]
\centering
{\includegraphics[width=7 cm, angle=0]{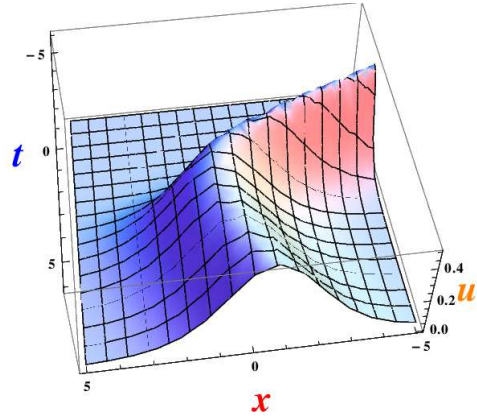}}

\caption{$x-t$ plot of the solution (\ref{accsolt}) with 
   $\alpha = 1$, $\delta=0.01$, $\epsilon= 0.0001$, $b_{0} =-20000 $,
$b_{1} =1$,$b_{2} =1$.  $x$ and $t$ are plotted in the 2 horizontal axes and $u(x,t)$ is plotted in the vertical axis.}
\label{Figure 3 }
\end{figure}

\begin{center}{\bf{ 2.4 ~~ C is a function of x only }}\end{center}

 The next case of interest deals with the situation where the leakage velocity is dependent on spatial  coordinate only. This has physical relevance  in practical situations where
the presence of porous medium in different parts of the bottom of the channel makes the leakage velocity also to be dependent on $x$.

From equations (\ref{bc1dl}), (\ref{bc2dl}) we obtain
\begin{equation}
 \eta_{t}+ f_{x}+ \epsilon f \eta_{x}+ \epsilon \eta f_{x}- \frac{1}{6}\delta^2 f_{xxx}
= C- \frac{1}{2}\delta^2 C_{xx}
\label{etat7} \end{equation}
\begin{equation}
 \eta_{x}+ f_{t}+ \epsilon f f_{x} - \frac{1}{2}\delta^2 f_{xxt}
= 0 \label{etax7}
\end{equation}

The function $f$ is expanded in a perturbation series as in earlier sections and the following choice is made for $f^{(0)}$,
\begin{equation}
 f^{(0)}(x,t) = \eta(x,t) +  B(x)+ O(\epsilon \delta^2)
\end{equation}
 where $B(x) = \int C(x) dx $.

 The compatibility of
the 2 equations (\ref{etat7}) and (\ref{etax7}) leads to,

\begin{equation}
 f^{(1)}_{x} = - \frac{1}{2}(\eta_{x} \eta - B C)
\end{equation}

\begin{equation}
 f^{(2)}_{x} =  \frac{1}{3}(\eta_{xxx})- \frac{1}{6}C_{xx}
\end{equation}

 Substituting these values of $f^{(1)}_{x},  f^{(2)}_{x}$ in any of the equations (\ref{etat7}) and (\ref{etax7})  leads to a single evolution equation

\begin{equation}
{\bar\eta_{t}}+ {\bar\eta_{x}}+ \epsilon \frac{3}{2} {\bar\eta}{\bar\eta_{x}} + \frac{\delta^2}{6}\bar\eta_{xxx} =\frac{2 }{3}B_{x}
+ \frac{\epsilon}{6} B B_{x} - \frac{\delta^2}{18} B_{xxx}
\label{kdvcx}
\end{equation}
where $ \bar{\eta} = \eta + \frac{2}{3} B$ and the equation(\ref{kdvcx}) has the form of a forced KdV equation.

{\bf 2.4.1 ~~Nature of the Solution}

For the case when the vertical fluid velocity at the bottom is a function of spatial coordinate only, the nonlinear equation for water wave propagation  given by equation (\ref{kdvcx}) is obtained as an inhomogeneous  KdV equation. The right hand side of eq. (\ref{kdvcx})  has a space KdV like form with the time coordinate  replaced by the spatial coordinate (here $x$).
 The mathematical elegance of this  equation enables one to obtain its solitary wave solutions in a simple manner considering the following two different cases:

\noindent {\bf Case (a)  }

A very simple situation occurs if the right hand side of the  equation (\ref{kdvcx}) is  taken to be zero. Using the scaling ($X = x/\delta$, $B'=\epsilon B$), the inhomogeneous part of  equation (\ref{kdvcx})  reduces to the  following:
\begin{equation}
12 B'_{X} + 3 B' B'_{X} -  B'_{XXX} = 0
\label{spacekdv} \end{equation}
A one-soliton like solution of this space KdV equation is obtained as 
\begin{equation}
 B'(X) =(-12 ) sech^2[\sqrt{3}(X - X_{0}]
\end{equation}
and the corresponding leakage velocity is given by
\begin{equation}
  C'(X) = \frac{\partial B'}{\partial X} = (24 \sqrt{3}) sech^2[\sqrt{3} (X - X_{0})]
 tanh[\sqrt{3} (X - X_{0})]
\label{leakv} \end{equation}

This leads to the conclusion that if $B'$ has the  functional form that   satisfies  (\ref{spacekdv}),
then the  evolution equation for the waves would  be  the standard KdV equation. Hence its  solution will  also be  given by the  standard KdV solitary wave solution moving with constant velocity.
 Thus for those functional forms of $B'$ and the corresponding
leakage velocity,  the wave solution will be practically unaffected by the leakage.
Since the solution of this case is the solution
of standard KdV,  it is not being shown  explicitly. The leakage velocity given in equation (\ref{leakv}) is plotted as Figure 4.

\begin{figure}[!h]
\centering

{\includegraphics[width=7 cm, angle=0]{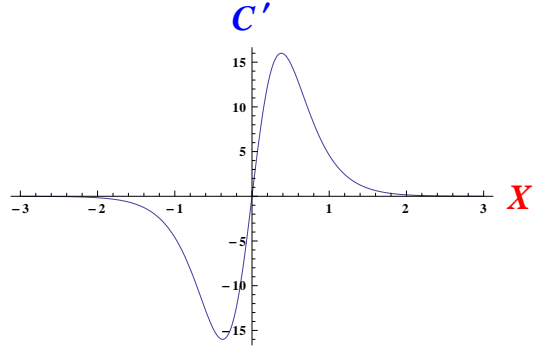}}

\caption{
Leakage velocity profile for $X_{0} = 0$ when $B'$ satisfies  (\ref{spacekdv}) }
\label{Figure 4}
\end{figure}

\noindent {\bf Case (b):  }

An interesting analytic solution of the equation (\ref{kdvcx})  arises if the leakage velocity function has a preassigned  form.
Considering the leakage velocity to be localized in space, (that is often consistent with certain physical condition), the following relevant form of $B(x)$ is chosen
  
\begin{equation}
 B (x) = m \tanh (n x),
\end{equation}
 where  $m$ and $n$ are  two  external small parameters dependent on the leakage profile.

Since $m, n$ are small parameters, in the following calculation  terms upto
$\sim m n $ will be retained and terms beyond  this order  are neglected. Hence second and third terms in r.h.s of the equation (\ref{kdvcx}) will produce higher order terms in $m,n$ and hence these are  neglected.

 Transforming to a new variable  $ u = \bar{\eta} - {2 B_{x} t}/{3}$,  the following equation is obtained from equation (\ref{kdvcx})
\begin{equation}
 u_{t}+ u_{x}+ \epsilon \frac{3}{2} u u_{x} + \frac{\delta^2}{6} u_{xxx} + \epsilon B_{x} u_{x} t = 0
\end{equation}

Further,  a transformation to a moving frame given by
$$ X = x -\frac{\epsilon B_{x}}{2} t^2 - t  ,  T=t $$
is carried out to obtain
\begin{equation}
 u_{T} + (\epsilon \frac{3}{2}) u u_{X} + (\frac{\delta^2}{6}) u_{XXX} = 0
\end{equation}
The above equation is finally cast in the standard form of a KdV equation by defining the following variables
$U = ({3 \epsilon}/{2 \delta^2}) u , T'= (\frac{\delta^2}{6}) T$
\begin{equation}
U_{T'} + 6 U U_{X} + U_{XXX} = 0
\label{kdvcx2} \end{equation}

The corresponding solitary wave solution  in the rest frame is given by
\begin{eqnarray}
 U(x,t) &=& \frac{\gamma}{2} Sech^2[\frac{\sqrt \gamma}{2}(x - \frac{\epsilon B_{x}}{2}t^2 -t( 1 + \frac{\gamma \delta^2}{6} )] \nonumber \\
&=& \frac{\gamma}{2}Sech^2[\frac{\sqrt \gamma}{2}(x -\frac{\epsilon m n Sech^2(n x)}{2} t^2 - t( 1 +\frac{\gamma \delta^2}{6} )],
\label{gensolx} \end{eqnarray}
where $\gamma$ is a constant.  The critical time $t_{c}$ at which the velocity of the wave becomes zero
is also a function of $x$. The 3D plot of the solution is shown in Figure (5).
\begin{figure}[!h]
\centering

{\includegraphics[width=7 cm, angle=0]{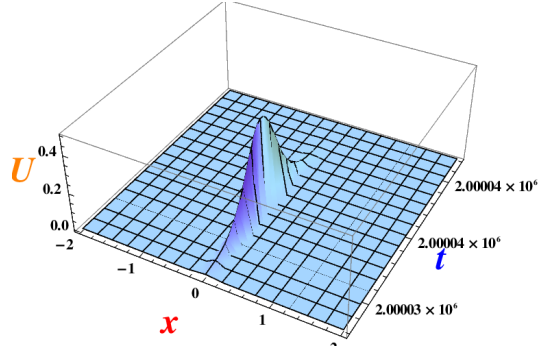}}
\caption{
$x-t$ plot of the  solution (\ref{gensolx})  with $ m = -0.1$, $n = 0.1$, $\gamma=1$ }
\label{Figure 5}
\end{figure}

\begin{center} {\bf {2.5~~  When $C$ is  function of both $x$ and $t$  }}\end{center}

Finally the problem will be treated in the most general manner by treating the leakage velocity $C$ as a function of both $x$ and $t$. The mathematical analysis is carried out in the following way.
As in all the earlier cases  $f$ is expanded  as 
\begin{equation}
 f = f^{(0)}+ \epsilon f^{(1)}+ \delta^2 f^{(2)}+ h.o.t,
\end{equation}
where $f^{(i)}$'s are functions of $\eta$ and its spatial derivatives.
 In the present case, the leakage velocity $C(x,t)$  is also expanded in a series of small parameters
in the following way 
\begin{equation}
 C(x,t) = \epsilon C_{1}(x,t) + \delta^2 C_{2}(x,t) + h.o.t
\end{equation}

Carrying out the same  kind of mathematical analysis as in the earlier cases the  following two equations
 are obtained from equations (\ref{bc1dl}) and (\ref{bc2dl}) 
\begin{equation}
 \eta_{t}+ f^{(0)}_{x}+ \epsilon (f^{(1)} _{x}+  \eta f^{(0)}_{x}+\eta_{x} f^{(0)} - C_{1})
+\delta^2 (f^{(2)}_{x}- \frac{1}{6}f^{(0)}_{xxx}-C_{2})= 0
\label{etat8}\end{equation}
\begin{equation}
 \eta_{x}+ f^{(0)}_{t}+ \epsilon (f^{(1)} _{t}+  f^{(0)} f^{(0)}_{x})
+\delta^2 (f^{(2)}_{t}- \frac{1}{2}f^{(0)}_{xxt})= 0
\label{etax8}\end{equation}
In order to make equations (\ref{etat8}), (\ref{etax8}) self-consistent as evolution equation for a 1d wave propagating to
the right, the following transformation is carried out

\begin{equation}
 f^{(0)}(x,t) = \eta(x,t) + h.o.t
\end{equation}
Since terms of the order of $\epsilon \delta^2$ are neglected, the following relations are obtained

$$ f^{(1)}_{t} \approx -\eta_{x}f_{\eta}^{(1)} =-f^{(1)}_{x} \nonumber $$ ,
$$ f^{(2)}_{t} \approx -\eta_{x}f_{\eta}^{(2)} =-f^{(2)}_{x} \nonumber $$
$$\eta_{xxt} \approx -\eta_{xxx} \nonumber $$

From the compatibility of the two equations (\ref{etat8}) and (\ref{etax8})  we get,
\begin{eqnarray}
 f^{(1)}_{x} &=& \frac{1}{2}(C_{1} - \eta_{x} \eta ) \nonumber \\
 f^{(2)}_{x} &=&  \frac{1}{3}(\eta_{xxx})+ \frac{1}{2}C_{2}
\end{eqnarray}

 Finally, a single evolution equation is obtained utilizing these functional forms of $f^{(1)}_{x},  f^{(2)}_{x}$ in any of the equations (\ref{etat8}), (\ref{etax8}) as:
\begin{equation}
\eta_{t} + \eta_{x} + (\epsilon \frac{3}{2}) \eta \eta_{x} +(\frac{\delta^2}{6})\eta_{xxx}  = \frac{1}{2}(\epsilon C_{1} + \delta^2 C_{2})
\label{kdvcxt1} \end{equation}
For the sake of mathematical simplicity the condition  $ C_{2}(x,t) = 0$ is assumed.  Since $\epsilon \approx \delta^2 $, it does not  break the generality of the problem. Using Galilean
transformation $\xi = x-t$ and $\tau = ({\delta^2}/{6}) t$ and defining the variable
$ u(\xi,\tau) =(\frac{3 \epsilon}{2 \delta^2}) \eta(\xi,\tau)$,  the final form of the evolution equation is obtained as,
\begin{equation}
 u_{\tau}+ 6 u u_{\xi} + u_{\xi \xi \xi} = \frac{9}{2} C_{1}(\xi,\tau)
\label{kdvfor} \end{equation}
The above equation has the form of a forced kdV equation.\\

{\bf 2.5.1 Nature of Solutions}\\

 For the sake of notational simplicity equation (\ref{kdvfor}) is expressed  in  variables $(x, t)$ as
\begin{equation}
u_{t}+ 6 u u_{x} + u_{xxx} = \frac{9}{2}\ C_{1}(x,t)
\label{forcedkdvxt} \end{equation}
In order to obtain  a solitary wave solution of the forced KdV equation,   the bilinearization 
technique\cite{kn:hirota_bl}  is used.

 Assuming
\begin{equation}
\frac{9}{2} C_{1}(x,t) = \frac{\partial D(x,t)}{\partial x}
\end{equation}
and using the bilinear transformation 

$$  u = 2 \frac{\partial^2}{\partial x^2}[\ln(F)], ~~D= \frac{G}{F}
$$
equation(\ref{forcedkdvxt}) transforms to the following bilinear equation
\begin{equation}
F F_{xt} - F_{x}F_{t}+ F F_{xxxx} + 3 F_{xx}^2 - 4 F_{x} F_{xxx}  = G F
\label{Feq} \end{equation}
{\bf In the previous cases, we have attempted to obtain the final solutions in the
form of solitary waves where phase gets modulated and amplitude remains constant.
In order to obtain similar solitary wave solutions in this case also
 the following   choice for the functions $G$ and $F$ have to be made,}
\begin{equation}
 G = h(t) Sech[x - v t -p(t)]
\end{equation}

\begin{equation}
 F = h_{1}(t)( \exp[x - v t -q(t)] +\exp[-x + v t + q(t)]),
\end{equation}
where $h(t), h_1(t), p(t)$ and  $q(t)$ are all arbitrary functions of time.
Substituting the expressions for $F$ and $G$ in equation (\ref{Feq}) leads to
\begin{equation}
 p(t) = q(t), ~~ h_{1}(t) = - \frac{h(t)}{2(-4 + v+ \frac{\partial p}{\partial t})}
\end{equation}
Thus, the analytic solution of the  forced KdV equation (\ref{forcedkdvxt})  is obtained as
\begin{equation}
 u = 2( Sech[x - v t -p(t)] )^2 \label{solxt}
\end{equation}
 with the forcing term $D$ given by
\begin{equation}
 D(x,t) = -(-4 + v + \frac{\partial q}{\partial t})( Sech[x - v t -p(t)] )^2
\end{equation}
Hence the leakage velocity $C_1 (x,t)$ is obtained in the form 
\begin{equation}
 C_{1}(x,t) = -2(-4 + v + \frac{\partial p}{\partial t})( Sech[x - v t -p(t)] )^2 \tanh[x - v t -p(t)]
\end{equation}
Since there is an arbitrary function $p(t)$  in the leakage velocity,  as well as in the solution, one can observe different types of waves excited by different forcing sources, i.e.  different functional forms of  $ p(t)$.  Here also the amplitude of the wave solution remains constant.

\begin{figure}[!h]
\centering
{\includegraphics[width=7 cm, angle=0]{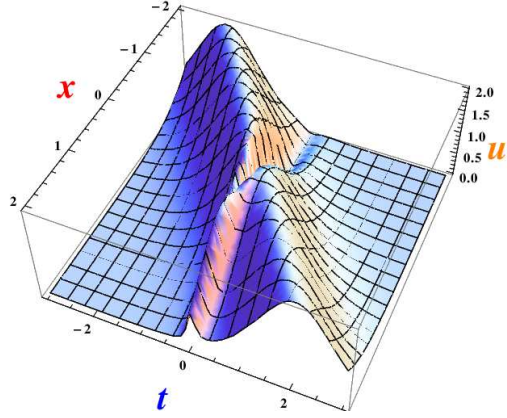}}

\caption{$x-t$ plot of the solution (\ref{solxt})  for $ q(t) = Sech(10 t) , v=1 $}

\label{Figure 6}
\end{figure}

\begin{center} {\bf Conclusive remarks}\end{center}

The problem of shallow water, unidirectional, weakly nonlinear,  surface wave propagation in a water channel 
is treated (in a way distinct from existing literature) by incorporating all the information related to the interactions of water and the bed  in the form of a non-trivial  penetration condition. When the vertical fluid velocity at the bottom  is constant then it is shown that the evolution equation is given by a forced KdV equation with a constant forcing term. The solitary wave solutions of such equation  will contain a constant retardation term in the argument of the function, while the amplitude will remain constant. When the leakage  is a function of time,   a time dependent  retardation   appears in the argument of the solution while the amplitude still remains constant.
When the leakage velocity is  only space dependent, two different kinds of solitary wave solutions are obtained analytically. For those functional forms of B(x), satisfying the stated
space KdV like equation, the solution will remain unaffected by the leakage and becomes identical to the standard KdV soliton with constant velocity.
 When the leakage velocity is a slowly varying function localized in space a  soliton solution with constant amplitude is obtained under certain approximation related to the slowness  of  variation of the leakage velocity profile. When the vertical fluid velocity  at the bed is assumed to be function of both space and time, the  bilinearization technique has been employed to obtain the  solutions to the evolution equation.  The technique yields a
 solitary wave solution  with a  constant amplitude   and an arbitrary function of time appearing in the argument of the solution as well as in the argument of the leakage velocity profile. The nature of the solution can be modulated by choosing  different forms of this arbitrary function.

The present work can be extended to address certain interesting problems.    Retaining the higher powers
of the expansion parameters $\epsilon$, $\delta ^2$, other evolution equations will be obtained
that are applicable to the other portions of the channel. Also, in the present work, the channel depth is  assumed to be constant. If the variation of bathymetry is considered, then solving the problem analytically
in presence of leakage velocity,  is a challenging task. If transverse directions are involved then instead of KdV equation, multidimensional  type equations will evolve  which are nearer to the actual real physical situations.

\begin{center} {\bf Acknowledgments} \end{center}

The authors are deeply indebted to their Professor  Anjan Kundu for suggesting this problem,
and giving all the motivation and encouragement to see it completed.

\end{document}